\documentclass[a4paper,12pt]{article} 

\usepackage{amsmath}
\usepackage{color}
\usepackage{amssymb}
\usepackage{bm}
\usepackage{graphicx}
\begin{document}

\thispagestyle{empty}
\vspace*{-15mm}
{\bf OCHA-PP-353}\\
\vspace{10mm}
\begin{center}
{\Large\bf
Quantization Viewed as Galois Extension\\
}
\vspace{7mm}

\baselineskip 18pt

\vspace{2mm}
Mamoru Sugamoto${}^{1}$ and Akio Sugamoto${}^{2, 3}$, 

\vspace{2mm}

{\it ${}^{1}$  Research and Development Division, Apprhythm Co., 4-5-18 Honmachi, Chuo-ku, \\
Osaka 541-0053, Japan\\
${}^{2}$Tokyo Bunkyo Study Center, The Open University of Japan (OUJ), \\
Tokyo 112-0012, Japan \\
${}^{3}$Ochanomizu University, 2-1-1 Ohtsuka, Bunkyo-ku, Tokyo 112-8610, Japan}

\end{center}

\vspace{10mm}
\begin{center}
\begin{minipage}{14cm}
\baselineskip 16pt
\noindent
\begin{abstract}
Quantization is studied from a viewpoint of field extension.\\
If the dynamical fields and their action have a periodicity, the space of wave functions should be algebraically extended {\it \`a la} Galois, so that it may be consistent with the periodicity. This was pointed out by Y. Nambu three decades ago.  

Having chosen quantum mechanics (one dimensional field theory), this paper shows that a different Galois extension gives a different quantization scheme.  A new scheme of quantization appears when the invariance under Galois group is imposed as a physical state condition.  Then, the normalization condition appears as a sum over the product of more than three wave functions, each of which is given for a different root adjoined by the field extension.
\end{abstract}

\end{minipage}
\end{center}

\newpage

\section{Introduction}
In a macroscopic world the dynamics is controlled by classical mechanics, while in a microscopic world, it is described by quantum mechanics.  
A classical particle becomes a wave in quantum mechanics.  Waves propagating along different paths may interfere, resulting to be strengthened or weakened.
The path integral quantization manifestly represents this wave interference \cite{Feynman}.  A wave function $\psi(q, t)$ in quantum mechanics gives the ``probability amplitude'' of a particle located at position $q$ and time $t$.  Then, the transition of a wave function $\psi(q_0, 0)$ at time $0$ to that $\psi(q, t)$ at time $t$ can be described by a transition amplitude $U(q, q_0; t, 0)$, 
\begin{eqnarray}
\psi(q, t) = U(q, q_0; t, 0) \psi(q_0, 0),
\end{eqnarray}
where the transition amplitude is given in terms of the path integral \cite{Feynman}, 
\begin{eqnarray}
U(q, q_0; t, 0)=\int_{t'=0}^{t'=t} \mathcal{D} q(t') e^{\frac{i}{\hbar} S}.
\end{eqnarray}
Here, $\hbar$ is the Planck constant divided by $2\pi$, and $S$ is the action of the system, given by
\begin{eqnarray}
S=\int_0^{t} dt' L(q, \dot{q})= \int_0^{t} dt' \left(p\dot{q}-H(q, p)\right)=\int_{q_0}^q pdq' -\int_{0}^{t}dt' H(q, p),  \label{action integral}
\end{eqnarray}
where $p=\partial L/\partial \dot{q}$ and $L$ is the Lagrangian and $H$ is the Hamiltonian of the system. 
The word of partition function $Z$ is usually used for the transition amplitude between the infinite time interval,
\begin{eqnarray}
Z =\int_{t=-\infty}^{t=+\infty} \mathcal{D} q(t) e^{\frac{i}{\hbar} S},
\end{eqnarray}
but here we use the word also for the transition amplitude $U(q, q_0; t, 0)$ for a finite time interval.

Path integral means the integration (sum) over all the possible paths connecting $q_0$ with $q$ during time interval $t$.  Assume that only two paths, $P_1$ and $P_2$, are concerned, such as a beam starting from a source at $q_0$ is separated by two slits and ends at $q$ on the screen, then
\begin{eqnarray}
U(q, q_0; t, 0)=e^{\frac{i}{\hbar} S[P_1]}+e^{\frac{i}{\hbar} S[P_2]},
\end{eqnarray}
where $S[P_1]$ and $S[P_2]$ are obtained by Eq.(\ref{action integral}) along the paths $P_1$ and $P_2$, respectively.  This shows that $S/\hbar$ is the phase $\Phi$ of the wave, so that if the phase difference is $\Delta \Phi=2\pi m$, two waves are strengthened, but if it is $\Delta \Phi=2\pi (m+1/2)$, these waves are weakened ($m=$integer); this is the interference of waves:
\begin{eqnarray}
\Delta S=\hbar \Delta \Phi= \begin{cases} 
~2\pi \hbar m= hm &(\mathrm{strengthened}), \\
~2\pi \hbar (m+1/2)=h(m+1/2) &(\mathrm{weakened}).
\end{cases}
\end{eqnarray}

Thus the quantization can be considered as the requirement of an ``additive periodicity" with a unit of $h$ (Planck constant) on the action $S$, namely
\begin{eqnarray}
S+ hm = S~~(m=\mathrm{integer, ~``additive~periodicity''}).
\end{eqnarray}

This ``additive periodicity'' in the power of the exponential implies the ``multiplicative periodicity'' in the base number of the exponential, that is 
\begin{eqnarray}
\left(e^{i 2\pi}\right)^m=1~~(m=\mathrm{integer, ~``multiplicative~periodicity''}).
\end{eqnarray}

At this point we may notice that the quantization procedure strongly depends on the periodicity of a circle.  If another periodicity is adopted, we may arrive at the different quantization procedure.  The periodicity controls the interference pattern, so that a different interference pattern may induce a different way of quantization.

Nambu afforded this viewpoint in 1987 \cite{Nambu}; he considered space, field and partition function as finite sets, $\mathbb{Z}_l, \mathbb{Z}_k$ and $\mathbb{Z}_h$, respectively, with $(l, k, h=\mathrm{integers})$.  The $\mathbb{Z}_m$ is a set of integers obtained by (mod $m$), $\mathbb{Z}_m=\{0, 1, 2, \cdots, m-1; (\mathrm{mod}~m) \}$, which means the ``charactereristic'' is $m$, that is for $\forall a \in \mathbb{Z}_m, ma=0 ~(\mathrm{mod}~ m)$.  Hence, $m$ is an additive periodicity.  However, we have to exponentiate the space variables to 
obtain the field (this is a procedure to make a classical wave), and then exponentiate the action again to obtain the partition function (this second stage of exponentiation gives the quantization procedure).  

Nambu concentrated on the first stage exponentiation to make classical waves, since his target was to find the Poincar\'e cycle (the recurrence time) \cite{Poincare} of the heat equation and the wave equation in classical mechanics.  

When we describe a classical wave as a function of space-time variables, we need the exponentiation.  Nambu considered this exponentiation as $(w_e)^t (z_p)^x$, where $(t, x)$ is time and space coordinates in $\mathbb{Z}_l$; $w_e$ and $z_p$ are elements in $\mathbb{Z}_k$ labeled by the indices $e$ and $p$, respectively.   The variety of the choice $(e, p)$ gives the variety of angular frequency (energy $e$ in the particle picture) and wave vector (momentum $p$ in the particle picture) of the wave.

The ``additive period'' of the space $\mathbb{Z}_l$ is $l$, but the ``multiplicative period'' of the field in $\mathbb{Z}_k$ is not simple, but a more delicate, reflecting the integer property of $k$.  To find this multiplicative period, Euler's theorem is useful \cite{Euler}, \cite{textbook on Galois}, which states that if $a$ and $n$ are relatively prime, that is, the greatest common divisor (gcd) of $a$ and $n$ is 1, gcd$(a, n)=1$, then $a^{\varphi(n)}=1~(\mathrm{mod}~n)$.  Here $\varphi(n)$ is the Euler's totient function. For $n=p_1^{a_1}p_2^{a_2} \cdots$ ($p_i$ are prime numbers and $a_i$ are integers), we have
\begin{eqnarray}
\varphi(n)=n\left(1-\frac{1}{p_1}\right)\left(1-\frac{1}{p_2}\right)\cdots.
\end{eqnarray} 

Now, we can easily find a multiplicative period for $\mathbb{Z}_k$.  The multiplicatively closed set in $\mathbb{Z}_k$ is denoted by $\mathbb{Z}^{\times}_k$, and is constructed of all the elements relatively prime to $k$, so that the order (the number of elements) of $\mathbb{Z}^{\times}_k$ is $\varphi(k)$, and $\forall a \in \mathbb{Z}^{\times}_k$, the following multiplicative periodicity holds:
\begin{eqnarray}
a^{\varphi(k)}=1~~(\mathrm{mod}~ k). 
\end{eqnarray}

If the proper matching between the multiplicative period for the field and the additive period in space-time fails, such as $\varphi(k) \ne l $, then we have to adjoin to $\mathbb{Z}_k$ the roots of the $l$-th algebraic equation with coefficients in $\mathbb{Z}_k$
\begin{eqnarray}
f(X)=X^l-1=0 ~(\mathrm{mod}~k).
\end{eqnarray}
The field so obtained was named $\mathbb{Z}_{k,~l}$ by Nambu.  If $k$ is prime, then $\mathbb{Z}_{k,~l}$ becomes a field in mathematics.  Accordingly, $\mathbb{Z}_{k,~l}$ is the Galois extension of the field $\mathbb{Z}_{k}$, by adjoining the roots of $f(X)=0$ \cite{Galois}, \cite{textbook on Galois}.  Now the matching of the ``muliplicative period'' and the ``additive period'' $l$ is guaranteed by the nontrivial roots $\{\zeta_l, (\zeta_l)^2, \cdots,(\zeta_l)^{l-1}\}$, since $(\zeta_l)^l=1$ holds for $\zeta_l=e^{2\pi i/l}$.

Another Galois extension is required, when we perform the second stage exponentiation to obtain the the wave function from the classical action.  Nambu said this part is the most radical one, and gave a rough sketch, but had not fully analyzed it.

Accordingly, the purpose of this paper is to pursuit this ``second step exponentiation'',  or the``quantization'', being required to derive quantum wave function function from classical action.  As a result, a different quantization scheme appears from a different Galois extension.

\vspace{5mm}
In the next section, classical mechanics on discrete fields is examined.  In Sec. 3, its path integral quantization is studied.  Transformation of wave function by Galois group is examined in Sec. 4.  Some examples are given in Sec. 5. The final section is devoted to conclusion and discussion. 

\section{Classical mechanics of discrete fields with discrete time steps}
The purpose of this paper is to study the quantization procedure of a model in which field variables take discrete values on a discrete parameter space (discrete time).  We begin with the classical mechanics, before examining the quantization in the subsequent sections. 

Let us define three kinds of space; the (parameter) space $\mathcal{M}$, the space of field variables $\mathcal{F}$, and the space of wave functions (or partition functions) $\mathcal{Z}$.  (Note that Nambu's original choice is $\mathcal{M}=\mathbb{Z}_l, ~\mathcal{F}=\mathbb{Z}_k$, and~$\mathcal{Z}=\mathbb{Z}_h$.)

We are going to study a simple quantum mechanics in which the parameter space is one dimensional space of time $t$; this space is $\mathcal{M}=\mathbb{Z}$ to which $t$ belongs.  The field variable is a generalized coordinate $q(t)$, and is assumed to belong to a finite space; our choice of this space is $\mathcal{F}=\mathbb{Z}_n =\mathbb{Z}/n\mathbb{Z} \ni q$ ($n=$ integer) to which one component field $q$ belongs.  These $\mathcal{M}$ and $\mathcal{F}$ are enough to discuss the classical motion or the classical path.  However, when we consider the quantum mechanics, partition functions or wave functions are inevitable, and we have to know what the space $\mathcal{Z}$ be; it will be discussed in the subsequent sections in relation to what kind of Galois extension is appropriate for $\mathcal{Z}$.  We consider $\mathcal{Z}$ before Galois extension be any field $K$.  In the examples, however, the quotient number field $\mathbb{Q}$ is taken, since no multiplicative periodicity exists in $\mathbb{Q}$ while the space of the classical fields $\mathcal{F}$ has an additive period of $n$, giving a manifest discrepancy between periods.

The Lagrangian $L(t)$ is defined by
\begin{eqnarray}
L(t)=\{q(t)-q(t-1)\}^2-V(q(t)), 
\end{eqnarray}
with a potential $V(q)=a_0 + a_1 q + a_2 q^2 +\cdots \in \mathcal{F}[q]$; $\mathcal{F}[q]$ is a standard notation of polynomial of $q$ with coefficients in $\mathcal{F}$.  The coefficients are chosen in $\mathcal{F}=\mathbb{Z}_n$, so that rational numbers such as $1/2$, $1/3$, {\it e.t.c.} do not appear, unless $n = \mathrm{prime}$.  We keep $n$ be any integer, but sometimes feel it more convenient to take $n$ be prime number.

The action $S$ between $t=0$ and $t$ ($t>0$) is defined by
\begin{eqnarray}
S(t, 0)= L(1)+ \cdots + L(t) =\sum_{i=1}^{t} L(t).
\end{eqnarray}
Following Euler \cite{Euler variational method}, the variational principle states that the action is invariant if $q$ is changed ``infinitesimally'' at any single point $i$ $(0< i <t)$ from $q(t)$ to $q'(t)$:
\begin{eqnarray}
0=\delta S \equiv S'(t, 0)-S(t, 0) &=& -\{q'(i)-q(i)\} \times \left\{ 2(q(i+1)-q(i)-q'(i)+q(i-1)) \right. \nonumber \\
&&+ a_1+ a_2(q(i)+q'(i)) + a_3 (q(i)^2 +q(i)q'(i)+q'(i)^2) \nonumber \\
&& \left. + a_4 (q(i)^3 +q(i)^2q'(i)+q(i)q'(i)^2+q'(i)^3) +\cdots \right\}. \label{variation of action} 
\end{eqnarray}

We have assumed that $q$ belongs to $\mathcal{F}=\mathbb{Z}_n$, so that the minimum variation of $q$ is $\pm 1$.  When we apply the variational principle, we have to take the limit of $q' \to q$. We consider here the following ``limit'' as the infinitesimal variation:
\begin{eqnarray}
0=\lim_{q' \to q} \left[ \frac{\delta S}{\delta q(t)} \right]= \lim_{q' \to q} \left[\frac{S'(t, 0)-S(t, 0)}{q'(i)-q(i)}\right], 
\end{eqnarray}
then the following equation appears for any $i$ ($0<i<t$):
\begin{eqnarray}
 2 ~\ddot{q}(i)= - \frac{\partial V}{\partial q}(i),
\end{eqnarray}
where the discrete version of derivatives are defined by
\begin{eqnarray}
\ddot{q}(i)&\equiv& q(i+1)-2q(i)+q(i-1), \label{the 2nd derivative}\\
\frac{\partial V}{\partial q}&\equiv& a_1+ 2a_2q + 3a_3 q^2 + 4a_4 q^3 +\cdots. \label{discrete derivatives}
\end{eqnarray}
In this way, the Newton-Euler-Lagrange equation of motion is ``derived'', which is considered to give the classical motion of a particle under the influence of the potential $V(q)$.  

Next we examine the conservation of energy.  It is natural to define the energy $E$ by
\begin{eqnarray}
E(t) \equiv \{q(t)-q(t-1)\}^2+V(q(t)).
\end{eqnarray}
Then, the temporal change of energy, $E(t+1)-E(t)$, can be estimated as follows:
\begin{eqnarray}
E(t+1)-E(t)&=&\{q(t+1)-q(t-1)\} \times \left\{ q(t+1)-2q(t)+q(t-1)\} + \{q(t+1)-q(t) \}  \right. \nonumber \\
&&\times \left\{ a_1+ a_2(q(t)+q(t+1)) + a_3 (q(t)^2 +q(t)q(t+1)+q(t+1)^2) \right. \nonumber \\
&& \left. + a_4 (q(t)^3 +q(t)^2q(t+1)+q(t)q(t+1)^2+q(t+1)^3) +\cdots \right\}. \label{energy difference}
\end{eqnarray}

By comparison of Eq.(\ref{variation of action}) and Eq.(\ref{energy difference}), we can find a delicate difference in the right hand sides; 
the role of $q'(t)$ in Eq.(\ref{variation of action}) is played by $q(t+1)$ in Eq.(\ref{energy difference}).

This means that if the time development is not rapid, and $q(t+1) \approx q(t)$, then Newton-Euler-Lagrange equation determines the conservation of energy, that is, the conservation of energy holds even in this discrete case, 
\begin{eqnarray}
\partial_t E(t) = \{q(t+1)-q(t) \} \times \left\{ 2 \ddot q(t) + \frac{\partial V(q)}{\partial q}(t) \right\}=0. 
\end{eqnarray}
In general, the equation of motion and the conservation of energy differs delicately.

Hamilton's equation of motion can also be written as
\begin{eqnarray}
&&4 \dot q(t)= \frac{\partial H(q, p)}{\partial p}(t), \\
&&\dot p(t)=-\frac{\partial H(q, p)}{\partial q}(t),
\end{eqnarray}
where the momentum is defined by $p(t)\equiv 2(q(t+1)-q(t))$ and is an even integer, so that the Hamiltonian is given by $H(q, p)=\frac{1}{4}p^2+V(q)$.

Consider $E$ be constant, then the energy conservation implies
\begin{eqnarray}
p(t)^2 =4\left[ E-V(q(t))\right].
\end{eqnarray}
If $q(t)$ belongs to $\mathbb{Z}_n$, then the momentum $p(t)\equiv 2(q(t+1)-q(t))$ also belongs to $\mathbb{Z}_n$.  So do the parameters of the potential $a_0, a_1, \cdots$.  Therefore, the classical trajectory depicts a discrete curve $(E)$ as a set of the points $\{(q(t), p(t))\in (\mathbb{Z}_n, \mathbb{Z}_n) \vert t=0, 1, 2, \cdots \}$.  The curve $(E)$ so obtained looks like an elliptic curve \cite{elliptic curve}:
\begin{eqnarray}
(E):~y^2=4E-4\left(a_0+a_1 x+a_2 x^2 + a_3 x^3 + a_4 x^4 +\cdots\right), \label{classical trajectory}
\end{eqnarray}
if the phase space $(q, p)$ is denoted by $(x, y)$.  See Sec. 6 for a further discussion on the classical trajectories.

\section{Path integral formulation of quantum mechanics}
In order to construct the path integral formulation of quantum mechanics for the model given in the last section, the factor $e^{i/\hbar}$ appeared in the usual path integral should be replaced by an element in $\mathcal{Z}$.  Choosing a proper element $g$ in $\mathcal{Z}$, we will perform the following replacement:
\begin{eqnarray}
e^{\frac{i}{\hbar}S(t, 0)}  \rightarrow g^{S(t, 0)}.
\end{eqnarray}
There is a variety in the choice of $g$, corresponding to the variety of waves, or that of interference pattern of waves; the choice of $g$ is by no means unique, that will be discussed in the next section. 

Then, the path integral expression of the wave function for our model reads
\begin{eqnarray}
\psi(q, t)= \sum_{\mathrm{path}P(q_0 \to q)} g^{S(t, 0; P)}\psi(q_0, 0)=\sum_{q(1)} \cdots \sum_{q(t-1)} g^{L(1)+ \cdots + L(t)}  \psi(q_0, 0), \label{path integral expression of psi}
\end{eqnarray}
where $q(1), \cdots, q(t-1)$ in the sum run over all the elements in $\mathcal{F}=\mathbb{Z}_n$, which generates the sum over all the paths $P(q_0 \to q)$, connecting two fixed end points $q_0$ and $q$ and passing through intermediate points $\{q(1), \cdots, q(t-1)\}$.  Here $S(t, 0; P)$ is the action for a given path $P$.\footnote{We have not introduced a factor $1/A$ in defining the sum over $q$.  Therefore the definition of wave function differs from that of Feynman \cite{Feynman}.}

Then, we have
\begin{eqnarray}
\psi(q, t+1)= \sum_{q'} g^{L(q)}\psi(q', t)=\sum_{q'} g^{(q-q')^2-V(q)} \psi(q', t) =g^{-V(q)} \sum_{\xi} g^{\xi^2} \psi(q-\xi, t).
\end{eqnarray}
Here, $\xi=q-q'$, and the sum over $q'$ becomes the sum over $\xi$ in the space $\mathbb{Z}_n$. 

As for $\psi(q, t-1)$, it is obtained by replacing $g \to g^{-1}$ in $\psi(q, t+1)$, namely
\begin{eqnarray}
\psi(q, t-1)=\sum_{q'} g^{-L(q)}\psi(q', t)=\sum_{q'} (g^{-1})^{(q-q')^2-V(q)} \psi(q', t) =g^{V(q)}  \sum_{\xi} (g^{-1})^{\xi^2} \psi(q-\xi, t).~~
\end{eqnarray}

Thus, we obtain the temporal development in our model:
\begin{eqnarray}
\frac{1}{2}\{\psi(q, t+1)-\psi(q, t-1)\}=\frac{1}{2}\sum_{\xi} \left\{g^{\xi^2-V(q)}-(g^{-1})^{\xi^2-V(q)}\right\} \psi(q-\xi, t).
\end{eqnarray}
The left hand side is the time derivative in discrete case $\partial_t \psi(t, q) \equiv \frac{1}{2} \{\psi(t+1, q)-\psi(t-1, q)\}$, so that the temporal development of our system is described by
\begin{eqnarray}
\partial_t \psi(t, q) =\hat{H} \psi(q, t),
\end{eqnarray}
where the Hamiltonian operator $\hat{H}$ is defined by
\begin{eqnarray}
\hat{H} \psi(q, t)=\frac{1}{2}\sum_{\xi} \left\{g^{\xi^2-V(q)}-(g^{-1})^{\xi^2-V(q)}\right\} \psi(q-\xi, t). \label{quantum Hamiltonian}
\end{eqnarray}

To obtain a Shr\"odinger like equation, Taylor series expansion in a finite variable $\xi$ in $\mathcal{F}=\mathbb{Z}_n$ is necessary:
\begin{eqnarray}
\psi(q-\xi)= \psi(q) - \xi \partial_q \psi(q) +  \frac{1}{2} \xi^2 \partial_q^2 \psi(q)+ \cdots,
\end{eqnarray}
where the discrete derivatives are given in Eq.(\ref{discrete derivatives}).\footnote{If we expand as $\psi(q+\xi)=\sum_{m=0}^{\varphi(n)} a_m \xi^m$, then the definition of discrete derivatives in (\ref{discrete derivatives}) determines $a_m=\frac{1}{m!}\partial_q^m \psi(q)$. This is, however, too naive. In reality we have to consider the $\mathbb{Z}_n$ property of $q$, by taking into account a number theoretical refinement. Without such refinement, the Taylor expansion may not work, unless $n=\infty$. }

Let us introduce the following sums:\footnote{These sums, $\{A_0(g), A_2(g), \cdots\}$, have surely important meanings in number theory.  For example, when $m=n=\mathrm{prime}$, $A_0(g)$ is the famous ``Gauss sum" $G(k, n)$, relevant to the quadratic quotient problem of integer \cite{Gauss};
$A(e^{2\pi i k/n})= G(k, n) \equiv \sum_{\xi=1}^{n} e^{2\pi i \frac{k \xi^2}{n}}$ which takes  $0$ for $n=2, ~\left(\frac{k}{n}\right)\sqrt{n}$ for $n\equiv 1 ~(\mathrm{mod}~4), ~\mathrm{and}~i\left(\frac{k}{n}\right)\sqrt{n}$ for $n\equiv 3~(\mathrm{mod}~4)$.  Here $\left(\frac{k}{m}\right)$ is the Legendre symbol.  It is 0 for $k=0$, but it takes $1$ for $k=a^2~(\mathrm{mod}~m)$, and $-1$ for $k \ne a^2~(\mathrm{mod}~m)$ with an integer $a$. See Sec.6.}
\begin{eqnarray}
A_0(g)=\sum_{\xi} g^{\xi^2}, ~A_1(g)=\sum_{\xi} g^{\xi^2} \xi= 0,~A_2(g)=\sum_{\xi}g^{\xi^2} \xi^2, ~A_3(g)=\sum_{\xi} g^{\xi^2} \xi^3= 0, \cdots.
\end{eqnarray}

Then, we obtain the following Shr\"odingier like equation,
\begin{eqnarray}
\partial_t ~\psi(q, t) =\left[V_Q(q)+ \frac{1}{2m_Q(q)} \partial_q^2 +\cdots \right] \psi(q, t),  \label{discrete Schroediger eqn}
\end{eqnarray}
where the ``quantum'' potential $V_Q(q)$ and the ``quantum mass'' $m_Q(q)$ are defined by
\begin{eqnarray}
V_Q(q)&\equiv&\frac{1}{2} \left( A_0(g) g^{-V(q)}- A_0(g^{-1}) g^{V(q)} \right),  \\
\frac{1}{m_Q(q)}&\equiv& \frac{1}{2} \left( A_2(g) g^{-V(q)}- A_2(g^{-1}) g^{V(q)}\right).
\end{eqnarray}
It is noted that the ``quantum potential and mass'' are not equal to the classical ones, but they depend on $g$ as well as on $q$.  In the above Sch\"odinger like equation, a familiar $i$ does not appear, since whether $i$ is included in $\mathcal{Z}$ or not depends on the choice of $g$.   

A stationary state, with an energy eigen-function $\psi_E(q, t)$ and an energy eigen-value $E$, can be determined by 
\begin{eqnarray}
\hat{H} \psi_E(q, t)=\frac{1}{2}\sum_{\xi} \left\{g^{\xi^2-V(q)}-(g^{-1})^{\xi^2-V(q)}\right\} \psi_E(q-\xi, t)=E \psi_E(q, t). \label{stationary states}
\end{eqnarray} 

It is important to note that the wave equation depends on the choice of $g$ in $\mathcal{Z}$ to which the partition function and the wave function belong.  This choice determines the quantization scheme, which depends on what kind of wave picture is considered in the quantum mechanics.  The choice of $g$ is related to the additive periodicity in the field space $\mathcal{F}$.  This point will be discussed in the next section.

\section{Transformation of wave function by Galois group}
As was stated in the introduction, we are going to consider the space $\mathcal{Z}$ of wave functions and partition functions be a finite Galois extension of a field $K$, by adjoining all the roots $\{\alpha_1, \alpha_2, \cdots, \alpha_m \}$ of $f(X)=0$ to $K$; $f(X)$ is a polynomial of the order $m$ having coefficients in $K$, that is, $f(X) \in K[X]$. 

 The $\mathcal{Z}$ is called a splitting field of a polynomial $f(X)$, and is expressed as $\mathcal{Z}=K[X]/(f(X))$; $(f(X))$ is a prime ideal of $K[X]$ defined by modulo $f(X)$. The field $K$ can be either $\mathbb{Z}_p$ ($p$ is a prime number) or the set of rational numbers $\mathbb{Q}$.  If we assume that the degree of the Galois extension, $[\mathcal{Z}: K]$, is finite $D$, then the order (the number of elements) of the corresponding Galois group $G$ is equal to $D$:
\begin{eqnarray}
G=Aut_K(\mathcal{Z})=Gal (\mathcal{Z}/K), ~~\mathrm{and}~~|G|=D,
\end{eqnarray}
where $Aut_K(\mathcal{Z})$ is a set of all the automorphisms of $\mathcal{Z}$, by keeping the elements in $K$ invariant.

In the following, we add a suffix $g$ to the symbol of wave function like $\psi_g(q, t)$ to specify the basis $g$ of powers explicitly, and write it as follows: 
\begin{eqnarray}
\psi_g(q, t)=\sum_{\mathrm{path}P(q_0 \to q)} g^{S(t, 0; P)}\psi_g(q_0, 0).
\end{eqnarray}

The field $K$ is extended to the field $\mathcal{Z}$, by adjoining all the roots $\{\alpha_1, \cdots, \alpha_m\}$ of the algebraic equation $f(X)=0$, so that the following elementary symmetric polynomials defined by the wave functions are invariant under any operation in $G$:
\begin{eqnarray}
G-\mathrm{invariants}=\begin{cases}
~\mathcal{S}_1(q, t) \equiv \sum_{i=1}^m \psi_{\alpha_i}(q, t), \\
~\mathcal{S}_2(q, t) \equiv \sum_{1 \le i < j \le m} \psi_{\alpha_i}(q, t) \psi_{\alpha_j}(q, t), \\
~\mathcal{S}_3(q, t) \equiv \sum_{1 \le i < j < k \le m} \psi_{\alpha_i}(q, t) \psi_{\alpha_j}(q, t)\psi_{\alpha_k}(q, t), \\
~\cdots \cdots \\
~\mathcal{S}_m(q, t) \equiv \psi_{\alpha_1}(q, t) \psi_{\alpha_2}(q, t)\cdots \psi_{\alpha_m}(q, t).
\end{cases}
\end{eqnarray}
This is easily understood, since any element in $G=\{\sigma_1, \sigma_2, \cdots, \sigma_D\}$ executes a permutation of the roots:
\begin{eqnarray}
&&\sigma_i (\alpha_j)= \alpha_{\sigma_i(j)}, ~\mathrm{and} \\
&&\sigma_i=\begin{pmatrix} 
1, &2, &\cdots, &m \\
\sigma_i(1), &\sigma_i (2), &\cdots, &\sigma_i (m) \\
\end{pmatrix}.
\end{eqnarray}

The time evolution discussed in the last section shows that the time reversal transformation $\mathcal{T}$ induces $g\to g^{-1}$, namely
\begin{eqnarray}
 \psi_g(q, t)^{\mathcal{T}}=\psi_g(q, -t)=\psi_{g^{-1}}(q, t).
 \end{eqnarray}
 
 Now let the $G$-invariant wave functions be physically acceptable, and examine these $G$-invariant wave functions, especially $\mathcal{S}_m(q, t)$.  To do this, introduce the elementary symmetric polynomials for the roots:
 \begin{eqnarray}
s_1 = \sum_{i} \alpha_i, 
~s_2 = \sum_{i < j} \alpha_i \alpha_j, 
~s_3 = \sum_{i < j < k} \alpha_i \alpha_j \alpha_k, 
~\cdots \cdots, 
~s_m = \alpha_1\alpha_2 \cdots \alpha_m.
\end{eqnarray}
 
Then, by using $\psi_g(q, t+1)= g^{-V(q)}\psi_g(q, t)$, we have
\begin{eqnarray}
\mathcal{S}_m (q, t+1) = (s_m) ^{-V(q)} \mathcal{S}_m(q, t).
\end{eqnarray}

Now we can understand that if $s_m=\alpha_1 \alpha_2 \dots \alpha_m=1$ holds, then the $q$ dependence in the coefficient disappears, and we have a time independent product.
\begin{eqnarray}
\sum_q \psi_{\alpha_1}(q, t+1)\psi_{\alpha_2}(q, t+1) \cdots \psi_{\alpha_m}(q, t+1)= \sum_q \psi_{\alpha_1}(q, t)\psi_{\alpha_2}(q, t) \cdots \psi_{\alpha_m}(q, t) .
\end{eqnarray}
 
Therefore, we can construct a quantum mechanics, based on the following normalization condition:
\begin{eqnarray}
\sum_q \psi_{\alpha_1}(q, t)\psi_{\alpha_2}(q, t) \cdots \psi_{\alpha_m}(q, t)= 1~(\mathrm{or}~0),
\end{eqnarray}
which means the normalization to 1 is possible unless the sum is zero.

This normalization condition opens a new possibility in quantum mechanics, if more than three roots have to be adjoined in the process of quantization. 

The normalization condition is related to the concept of the existence probability.  In the usual case $\int dq~ \psi(q, t)^{\dagger} \psi(q,t)=1$ indicates that the wave function describes the one particle having the existing probability one.  However, in our new normalization condition for $m \ge 3$, $m$ particles form something like a ``bound state'' and the probability of its ``bound state'' is normalized.

Accordingly, it is convenient to consider the wave function be a set of $m$ component ones, namely, $\Psi(q, t) = (\psi_{\alpha_1}(q, t), ~\psi_{\alpha_2}(q, t), ~ \cdots, ~ \psi_{\alpha_m}(q, t))$.  The normalization condition can be written in the Dirac way, as
\begin{eqnarray}
\langle 0 \vert \cdot \vert \psi \rangle_{\alpha_1}\vert \psi \rangle_{\alpha_2} \cdots \vert \psi \rangle_{\alpha_m}=1~(\mathrm{or}~0),
\end{eqnarray}
where $\vert 0 \rangle$ denotes the vacuum.
If the time reversal is applied for the first component, we have
\begin{eqnarray}
_{\alpha_1^{-1}} \langle \psi \vert \cdot \vert \psi \rangle_{\alpha_2}\vert \psi \rangle_{\alpha_2} \cdots \vert \psi \rangle_{\alpha_{m}}=1~(\mathrm{or}~0),
\end{eqnarray}
since the time reversal changes an incoming state to an outgoing state, that is, $\vert \psi \rangle_{g}^{\mathcal{T}}=_{g^{-1}} \langle \psi \vert$. 

The energy eigen-value $E$ of $\Psi$, describing a ``bound state", is the sum of energies of the constituent particles,
\begin{eqnarray}
\hat{H} \Psi(q, t) = E \Psi(q, t) ~\mathrm{with}~E= E_1+E_2+ \cdots +E_m,
\end{eqnarray}
where $E$ is invariant under the Galois group $G=Gal(\mathcal{Z}/K)$.  

Thus, we can find that the quantum mechanics can be constructed so that the energy, the wave function, and the normalization condition are all invariant under the Galois group $G$.  Physical states can be expressed in terms of Galois invariants, that is, the Galois invariance be the physical state condition.

Now, it is time to discuss what a polynomial $f[X]$ should be taken.  The purpose of adjoining roots $\{\alpha_1, \alpha_2, \cdots, \alpha_m\}$ of $f[X]=0$ is to guarantee the ``multiplicative periodicity'' for the wave function $\psi$ so that it may be consistent with the ``additive periodicty'' for the field $q$. The wave function $\psi$ belongs to $\mathcal{Z}=K(\alpha_1, \alpha_2, \cdots, \alpha_m)$, while the field $q$ belongs to $\mathcal{F}=\mathbb{Z}_n$.  (See the introduction for the statement by Nambu \cite{Nambu}.) 

First, let us remind of an ambiguity of the wave function in quantum mechanics.  In the usual case, a phase factor $e^{i\theta}$ of the wave function is ambiguous and can not be fixed, and hence two wave functions having differ phase factors are equivalent.  

Our case is a little different from the usual one.  However, from Eqs.(\ref{path integral expression of psi}) and (\ref{quantum Hamiltonian}), we can understand the following: The Lagrangian $L$ belongs to $\mathcal{F}=\mathbb{Z}_n$, and is identified by modulus $n$.  The wave function should be also identified by modulus $n$ in the Lagrangian.  Thus, the wave functions should fulfill the following ``multiplicative periodicity'', including the $\pm$ ambiguity in the wave functions:
\begin{eqnarray}
g^{n}=g^{-n}=\pm 1.
\end{eqnarray} 
In our case $g$ and $g^{-1}$ appear as a combination of $g^L-(g^{-1})^L$, so that the ambiguity should be the same for both $g$ and its inverse $g^{-1}$, which yields $\pm 1$.  

Then, both a wave function $\psi$ and its time derivative $\dot{\psi}$ become equivalent to themselves under the additive periodicity of the field or the Lagrangian.  If the field $K$ satisfies this condition, then we can choose such $g$ from $K$.   However, if there is no such $g$ in $K$, an example for this is $K=\mathbb{Q}$, we have to introduce $g$ into $K$ as a root of the algebraic equation $f[X]=X^m \pm 1=0$, where $m$ is a divisor of $n$.  This is the reason why Galois extension is required.

In order to have a proper normalization condition, the roots should satisfy $\alpha_1 \cdots \alpha_m=1$.  This requires the algebraic equation is 
\begin{eqnarray}
X^m= \begin{cases} -1~(\mathrm{for}~m=\mathrm{even}) \\
+1~(\mathrm{for}~m=\mathrm{odd}),
\end{cases}
\end{eqnarray}
and the periodicity of the wave function becomes
\begin{eqnarray}
\psi(q+m, t)=\begin{cases} -\psi(q, t)~(\mathrm{anti-periodic~for}~m=\mathrm{even}) \\
+\psi(q, t)~(\mathrm{periodic~for}~m=\mathrm{odd}).
\end{cases}
\end{eqnarray}

\section{Some Examples}

We consider a number of simple examples with small $n$, where $n$ is the modulus of the field space $\mathcal{F}=\mathbb{Z}_n$, and find explicitly eigen-values and eigen-functions in these cases. The algebraic equation we examine is $f[X]=X^m\pm 1$.  We use a index $i=1-m$ to label the wave functions for different roots $\{\alpha_1, \alpha_2, \cdots, \alpha_m\}$, while use a index $(j)=(1)-(n)$ to label the different eigen-modes $(j)=(1)-(n)$ for a root, then the wave function becomes a $m \times n$ matrix; $\left(\psi_{i,(j)}\right)$ with $i=1-m, ~j=1-n$.

\vspace{3mm}
$\bullet$ \underline{(Example 1)}: $n=m=2, ~f_1(X)=X^2+1$, $\mathcal{Z}=\mathbb{Q}[i]$, and $V(q)=q^2$. 
 
The two roots are $\alpha_1=i$ and $\alpha_2=-i$.  These roots are related by the complex conjugation. In such a case, since $(\hat{H}_g)^{\dagger}=\hat{H}_{g^{\dagger}}$, we have $E_{g^{\dagger}}=(E_{g})^{\dagger}$ and $\psi_{g^{\dagger}}=(\psi_{g})^{\dagger}$.  In the following, the $q$-dependence of the wave function is given as a column vector, $\psi=(\psi(q=1), \psi(q=2), \cdots, \psi(q=n))^T$.  Remember that the wave function is anti-periodic in this case, $\psi(q+2)=-\psi(q)$.

1) For $g=\alpha_1=i$, we have
\begin{eqnarray}
\left(E_1, \psi_1(q) \right)=\begin{cases} \left(E_{1, (1)}=0, ~\psi_{1, (1)} \propto (0, 1)^T \right) \\
 \left(E_{1, (2)}=-i, ~\psi_{1, (2)} \propto (-1, 1)^T \right).
\end{cases}
\end{eqnarray}

2) For $g=\alpha_2=-i$, we have
\begin{eqnarray}
\left(E_2, \psi_2(q) \right)=\begin{cases} \left(E_{2, (1)}=0, ~\psi_{1, (1)} \propto (0, 1)^T \right) \\
 \left(E_{2, (2)}=i, ~\psi_{2, (2)} \propto (-1, 1)^T \right).
\end{cases}
\end{eqnarray}

The total energy is $E=E_1+E_2$, so that $E=\{0, i, -i\}$. 

The normalization condition of the wave function becomes as usual:
\begin{eqnarray}
\sum_{q} \psi_1(q, t)_{(a)} \psi_2(q, t)_{(b)} = \sum_{q}  \psi_1(q, t)_{(b)}^{\dagger} \psi_1(q, t)_{(a)}=1~(\mathrm{or}~0),
\end{eqnarray}
where $(a)$ and $(b)$ label the eigen-modes. 

\vspace{3mm}
$\bullet$\underline{(Example 2)}:  $n=m=3, ~ f_2(X)=X^3-1$. $\mathcal{Z}=\mathbb{Q}[i, \sqrt{3}]$, and $V(q)=q^2$. 

Three roots of $f_2(X)=0$ are $\alpha_1=1, \alpha_2=\omega, \alpha_3=\omega^2$, where $\omega=e^{2\pi i/3}$.  Remember that the wave function is periodic in this case, $\psi(q+3)=+\psi(q)$.
 
The eigen-values and eigen-functions are obtained as follows:\\
1) For $g=\alpha_1=1$, the Hamiltonian is zero matrix, and 
\begin{eqnarray}
\left(E_1, \psi_1(q) \right)= \left(E_{1, (1)-(3)}=0, ~\psi_1 \propto (a, b, c)^T \right),\end{eqnarray}
where $(a, b, c)$ are any constants.

2) For $g=\alpha_2=\omega$, we obtain
\begin{eqnarray}
\left(E_2, \psi_2(q) \right) =\begin{cases} \left(E_{2, (1)}=0, ~\psi_{2, (1)} \propto (0, 0, 1)^T \right), \\
\left(E_{2, (2)}=-\frac{\sqrt{3}}{2}i, ~\psi_{2, (2)} \propto (1, 0, 1)^T \right),\\
\left(E_{2, (3)}=-\frac{\sqrt{3}}{2}i, ~\psi_{2, (3)} \propto (0, 1, -1)^T \right).
\end{cases}
\end{eqnarray}

3) For $g=\alpha_3=\omega^2=\omega^{-1}$, its Hamiltonian is the complex conjugate of that for $g=\omega$, so that engen-values and eigen-functions are the complex conjugates of those for $g=\omega$, giving 
\begin{eqnarray}
\left(E_3, \psi_3(q) \right) =\begin{cases} \left(E_{3, (1)}=0, ~\psi_{3, (1)} \propto (0, 0, 1)^T \right), \\
\left(E_{2, (2)}=+\frac{\sqrt{3}}{2}i, ~\psi_{2, (2)} \propto (1, 0, 1)^T \right),\\
\left(E_{2, (3)}=+\frac{\sqrt{3}}{2}i, ~\psi_{2, (3)} \propto (0, 1, -1)^T \right).
\end{cases}
\end{eqnarray}

In this case of $n=m=3$, the total energy takes values $E=\{0, \frac{\sqrt{3}}{2}i, -\frac{\sqrt{3}}{2}i \}$.

The normalization condition becomes 
\begin{eqnarray}
\sum_{q} \psi_1(q, t)_{(a)} \psi_2(q, t)_{(b)} \psi_3 (q, t)_{(c)} =1~(\mathrm{or}~0),
\end{eqnarray}
where $(a, b, c)$ takes one of three eigen-states $(1)-(3)$. The state is given by the product of three wave functions, each of which is associated with the three roots $\{1, \omega, ~\mathrm{and}~ \omega^2\}$ of the algebraic equation $f[X]=X^3-1=0$. The adjoining of three roots are required in order for the wave function have the ``multiplicative periodicity'' of the order three, which is consistent with the ``additive periodicity'' of the order three for the field space $\mathbb{Z}_3$.   

\vspace{5mm}

Next, we examine (Example 3) with $m=2, n=6$ and (Example 4) with $m=3, n=6$.
In these examples, the space of wave functions is a little large $6 \times 6$ matrix. First  we write $L_{q,q'}=S(q, q'; 1, 0)$, $(q, q')=\{1, \cdots, 6 ~(\mathrm{mod}~6)\}$.  For a simple case of $V(q)=q^2$, we have
\begin{eqnarray}
L_{q, q'}=\begin{pmatrix} 
-1 &0 &3 &2 &3 &0 \\
3 &2 &3 &0 &-1 &0 \\
1 &-2  &3  &-2 &1 &0 \\
-1 &0 &3 &2 &3 &0 \\
3 &2 &3 &0 &-1 &0 \\
1 &-2 &3 &-2 &1 &0 
\end{pmatrix}
\end{eqnarray}

Then, the $(q, q')$ component of the Hamiltonian becomes $\hat{H}_{q, q'}=\frac{1}{2} \left(g^{L_{q, q'}}-(g^{-1})^{L_{q, q'}}\right)$, and we can find eigen-values and eigen-functions of the Hamiltonian for a given $g$.

\vspace{3mm}
$\bullet$\underline{(Example 3)}: $m=2, n=6$ with $V(q)=q^2$.

1) For $g=\alpha_1=i$, we have
\begin{eqnarray}
\left(E_1, \psi_1(q) \right)=\begin{cases} \left(E_{1, (1)}=0, ~\psi_{1, (1)} \propto (0, 0, 0, 0, 0, 1)^T \right) \\
\left(E_{1, (2)}=0, ~\psi_{1, (2)} \propto (0, 1, 0, 0, 0, 0)^T \right) \\
\left(E_{1, (3)}=0, ~\psi_{1, (3)} \propto (0, 0, 0, 1, 0, 0)^T \right) \\
\left(E_{1, (4)}=0, ~\psi_{1, (4)} \propto (-1, 0, 0, 0, 1, 0)^T \right) \\
\left(E_{1, (5)}=\frac{-\sqrt{7}-3i}{2}, ~\psi_{1, (5)} \propto (\frac{-1+\sqrt{7}i}{4}, \frac{-1+\sqrt{7}i}{4}, 1, \frac{-1+\sqrt{7}i}{4}, \frac{-1+\sqrt{7}i}{4}, 1)^T \right) \\
\left(E_{1, (6)}=\frac{\sqrt{7}-3i}{2}, ~\psi_{1, (6)} \propto (\frac{-1-\sqrt{7}i}{4}, \frac{-1-\sqrt{7}i}{4}, 1, \frac{-1-\sqrt{7}i}{4}, \frac{-1-\sqrt{7}i}{4}, 1)^T \right)
\end{cases}
\end{eqnarray}

2) For $g=\alpha_2=-i$, the eigen-values and eigen-functions are the complex conjugate of those for $g=i$.  Therefore, $E_{2, (5)}=\frac{-\sqrt{7}+3i}{2}$ and $E_{2, (6)}=\frac{\sqrt{7}+3i}{2}$.

The reason why $\sqrt{7}$ appears in the eigen-values and the eigen-functions is that the algebraic equation $h(\lambda)=0$ which determines the eigen-values is,
\begin{eqnarray}
h(\lambda)=\lambda^4(\lambda^2+3\lambda+4),
\end{eqnarray}
where $E=i\lambda$.  To realize the stationary states rigorously with non-zero energy we have to adjoin the root $\sqrt{7}$.  However, without adjoining $\sqrt{7}$, the approximate eigen-modes can exist, by choosing $\sqrt{7} \approx \frac{8}{3}$.  The error is only one per cent, since $\sqrt{7}=2.645 \dots$ and $\frac{8}{3}=2.666 \dots$. Therefore, such approximation can be accepted in our model defined on a discrete lattice. 

The total energy takes seven values $E=\{0, \pm\sqrt{7}, \pm\frac{\sqrt{7}+3i}{2}, \pm\frac{\sqrt{7}-3i}{2} \}$.

The normalization condition is as usual and is the same to (Example 1).

\vspace{3mm}
$\bullet$\underline{(Example 4)}: $m=3, n=6$ with $V(q)=q^2$.

1) For $g=\alpha_1=1$, the Hamiltonian is a zero matrix, and 
\begin{eqnarray}
\left(E_1, \psi_1(q) \right)= \left(E_{1, (1)-(6)}=0, ~\psi_1 \propto (a_1, \cdots, a_6)^T \right),\end{eqnarray}
where $a_i~(i=1-6)$ are any constants. 

2) For $g=\alpha_2=\omega$, we obtain
\begin{eqnarray}
\left(E_{2}, \psi_2(q) \right) =\begin{cases} \left(E_{2, (1)}=0, ~\psi_{2, (1)} \propto (0, 0, 0, 0, 0, 1)^T \right), \\
\left(E_{2, (2)}=0, ~\psi_{2, (2)} \propto (0, 0, 1, 0, 0, 0)^T \right),\\
\left(E_{2, (3)}=0, ~\psi_{2, (3)} \propto (-1, 0, 0, 1, 0, 0)^T \right), \\
\left(E_{2, (4)}=0, ~\psi_{2, (4)} \propto (0, -1, 0, 0, 1, 0)^T \right), \\
\left(E_{2, (5)}=-2\sqrt{3}i, ~\psi_{2, (5)} \propto (-1, 0, 1, -1, 0, 1)^T \right),\\
\left(E_{2, (6)}=-2\sqrt{3}i, ~\psi_{2, (6)} \propto (-1, 1, 0, -1, 1, 0)^T \right)
\end{cases}
\end{eqnarray}

3) For $g=\alpha_3=\omega^2=\omega^{-1}$, its Hamiltonian is the complex conjugate of that for $g=\omega$, so that engen-values and eigen-functions are the complex conjugates of those for $g=\omega$. In the present case, $E_{3, (1)-(4)}=0$, and non-vanishing eigen-values are  $E_{3, (5)(6)}=+2\sqrt{3}i$, and eigen-functions are the same as for $g=\omega$.

The total energy takes three values $E=\{0, \pm2\sqrt{3}i \}$.

The normalization condition is the same as in (Example 2):
\begin{eqnarray}
\sum_{q} \psi_1(q, t)_{(a)} \psi_2(q, t)_{(b)} \psi_3 (q, t)_{(c)} =1~(\mathrm{or}~0),
\end{eqnarray}
where $(a, b, c)$ takes one of six eigen-states $(1)-(6)$.

\vspace{3mm}

We can examine the model in a naive way, by finding explicitly the eigen-values and eigen-functions for a given $m$, $n$ and $V(q)$.  However, the more systematic analysis may exist, by using, for example, the expansion in terms of Fourier modes on the discrete space.  See Sec. 6 for a further discussion.  When $(m, n)$ are any integers, satisfying $m$ is a divisor of $n$, or replacing $\mathbb{Q}$ by a finite group $\mathbb{Z}_p$ and other finite Galois fields, the deeper and more delicate understanding of number theory will be required.

\section{Conclusion and discussions}
\subsection{Conclusion}
This paper considers how the space of wave functions is extended in the procedure of quantization, so that it may be consistent with the additive periodicity in the field space.  If the field space is $\mathcal{F}=\mathbb{Z}_n$, and the space of the wave function is $\mathbb{Q}$, the algebraic equation $f[X]=X^m \mp 1=0$ ($m$ a divisor of $n$, and $\mp1=+1$ for $m=$even and $\mp1=-1$ for $m=$odd.) should be solved, and the roots $\{\alpha_1, \alpha_2, \cdots, \alpha_m\}$ should be adjoined to $\mathbb{Q}$, forming the space of the wave functions be $\mathcal{Z}=\mathbb{Q}(\alpha_1, \alpha_2, \cdots, \alpha_m)$.  The wave function is defined in terms of path integral for each root.  When the quantum theory should be invariant under the Galois group $G=Aut_\mathbb{Q}(\mathcal{Z})$, the $G$-invariant combination of wave functions becomes physically acceptable.  Among them the sum over $q$ of the product of $m$ wave functions, each of which is defined for each root, is temporally invariant, and hence it can be used for the normalization condition.  The cases with $m \ge 3$ give new normalization conditions.  Some simple examples are discussed for $(m, n)=(2, 2), (3, 3), (2, 6)$ and $(3, 6)$.

\subsection{(Discussion 1) Classical trajectories and Poincare cycle} 
Let us discuss on the classical trajectory (\ref{classical trajectory}) described by an elliptic curve. It is true that there exists an Abelian group $A(E)$ on the curve $(E)$. The integer point $\bm{x}_0$ at $t=0$ on $(E)$ moves dynamically to another integer point $\bm{x}(t)$ after  $t$ time steps.  By introducing a time development operator $\hat T$ (Liouville operator in classical mechanics) for a single step $t \to t+1$, we have
\begin{eqnarray}
 \bm{x}(t)=(\hat{T})^t \bm{x}_0.
 \end{eqnarray}
As was studied by Nambu \cite{Nambu} we can find various Poincar\'e cycles $P$s (recurrent cycles), $\hat{T}^P\bm{x}_0=\bm{x}_0$.  Each Poincare cycle $P$ gives the group structure $\mathbb{Z}_P$,
\begin{eqnarray}
\mathbb{Z}_{P (\bm{x}_0)}= \{T, T^2, \cdots, T^P=1 \} \bm{x}_0.
\end{eqnarray}
In our case the total number of points are finite $n^2$, so that after $n^2$ time steps, the point in phase space will come back to the starting point.

Therefore, we have
\begin{eqnarray}
A(E)= \sum_{[\bm{x}_0]} \oplus \mathbb{Z}_{P ([\bm{x}_0])}, \end{eqnarray} 
where $[\bm{x}_0]$ denotes the equivalent class, giving the same trajectory $\mathbb{Z}_{P (\bm{x}_0)}$.

This Abelian group looks similar to the Modell-Weil group \cite{elliptic curve}, but both are different.

The problem in mathematics is to find integer points $(x, y)$ on an elliptic curve depicted on the plane of $\mathbb{Q}^2$, while in our problem of classical mechanics, all the point on a curve (a discrete trajectory) are integer points from the beginning.  Nevertheless, this kind of dynamical study of classical trajectories is interesting.  A problem is to classify the group $A(E)$ in terms of the parameters in the potential $\{a_0, a_1, \cdots\}$, or of the shape of the potential $V(q)$. Change of the classical potential can trigger the change of the group structure, forming a kind of phase structure.
 
\subsection{(Discussion 2) Eigenvalue problem using Jacobi $\vartheta$ function and its analog in number theory}
 We are tempted to continue to persuit the eigenvalue problem a little more, by using the Fourier expansion. 

Consider for example, the case with $n=\infty, ~f(X)=X^m\mp1, ~\mathcal{Z}=\mathbb{Q}[\alpha_1, \cdots, \alpha_m]$, and $V(q)=a_2 q^2$, where  the coupling constant (the spring constant) $a_2 \in \mathbb{Z}$ is chosen arbitrary. 

In this case, the space of fields is $\mathbb{Z}$, and hence the familiar  Fourier expansion can be applied for $\psi(q, t)$. The roots of $f(X)=0$ read
\begin{eqnarray}
\alpha_k=e^{i k \theta}, ~\mathrm{with}~k=\{1, 2, \cdots, m\},
\end{eqnarray}
where $\theta=\frac{2\pi}{m} \left(1+ \frac{1}{4}(1 \mp 1)\right)$.

The Fourier transform of a function $\psi(q, t)$ with a discrete variable $q \in \mathbb{Z}$ to a function $\tilde{\psi} (p, t)$ with a continuous variable $p \in [-\pi, \pi]$ is given by
\begin{eqnarray}
\psi(q, t)&=&  \frac{1}{2 \pi} \int_{-\pi}^{\pi} dp ~e^{-i pq}~\tilde{\psi}(p, t), ~\mathrm{and} \\
\tilde{\psi}(p, t)&=& \sum_{q=-\infty}^{\infty} ~e^{+i pq}~\psi(q, t).
\end{eqnarray}

Then, the eigen-value equation (\ref{stationary states}) can be written in terms of the Fourier transformed wave function $\tilde{\psi}(p, t)$:
\begin{eqnarray}
\frac{1}{2\pi} \int_{-\pi}^{\pi}dp'~ \hat{G}(p, p')~\tilde{\psi}(p', t)=E~ \tilde{\psi}(p, t), \label{equation in the momentum space}
\end{eqnarray}
where 
\begin{eqnarray}
 \hat{G}(p, p')=\frac{1}{2}\left( \sum_{q \in \mathbb{Z}} \sum_{\xi \in \mathbb{Z}} e^{-i( k\theta a_2 q^2+(p-p')q)} e^{i(k \theta  \xi^2+p\xi)}- \sum_{q \in \mathbb{Z}} \sum_{\xi \in \mathbb{Z}} e^{i( k \theta a_2 q^2-(p-p')q)} e^{i(-\theta k \xi^2+p\xi)}\right).
\end{eqnarray}

Using Jacobi's elliptic function $\vartheta_3$, defined by a sum or by a product:
\begin{eqnarray}
\vartheta_3(z, \tau)&=& \sum_{n=-\infty}^{\infty} e^{2\pi i \left( \frac{1}{2} \tau n^2  + zn \right)}=1+2 \sum_{n=1}^{\infty} q^{n^2} \cos(2\pi zn) \\
&=& \prod_{n=1}^{\infty} \left(1-q^{2n} \right)\left(1+2q^{2n-1}\cos(2\pi zn) +q^{4n-2} \right),
\end{eqnarray}
where $\tau$ and $z~ \in \mathbb{C}$, and $q=e^{\pi i \tau}$. 

Then, using the definition by sum, we have
\begin{eqnarray}
\hat{G}(p, p')=\vartheta_3 \left(\frac{p-p'}{2\pi}, -\frac{k\theta a_2}{\pi}\right) \vartheta_3 \left(\frac{p'}{2\pi}, \frac{k \theta}{\pi}\right) - \vartheta_3 \left(\frac{p-p'}{2\pi}, \frac{k\theta a_2}{\pi}\right) \vartheta_3 \left(\frac{p'}{2\pi}, -\frac{k \theta}{\pi}\right).
\end{eqnarray}

It is difficult for the authors to find the general solutions of (\ref{equation in the momentum space}).  What we can do is to approximate $\theta_3 \approx  1+2 e^{2\pi i \tau} \cos(2\pi z)$. This is an approximation of restricting the range of $\xi=q-q'$ and $q$ to the nearest neighbors $\{0,\pm1\}$, or of ignoring higher order derivatives than $\partial_q^2$.

In case of finite $n$ in general, we have to use a discrete version of the Fourier expansion,
\begin{eqnarray}
\psi(q, t)&=&  \frac{1}{n} \sum_{p=1}^{n} ~ e^{-2\pi i pq/n}~\tilde{\psi}(p, t), ~\mathrm{and} \\
\tilde{\psi}(p, t)&=& \sum_{q=1}^{n} ~e^{+2\pi i pq/n}~\psi(q, t).
\end{eqnarray}

Then, we need to know a discrete version $\Theta_3$ of the Jacobi's $\vartheta_3$ function:
\begin{eqnarray}
\Theta_3(z, \tau) \equiv \sum_{s=1}^{n} e^{2\pi i \left( \frac{1}{2} \tau s^2  + zs \right)/n},
\end{eqnarray}
defined by the sum or by the product.
This is a generalized Gauss sum which accomodates all the necessary quantities $A_0(g), A_2(g), \dots$ discussed in Sec 3. For $z=0$, $G(\tau n, n)$ is the original Gauss sum \cite{Gauss}. 
If we introduce the quartic coupling $a_4 q^4$ in the potential, the sum\begin{eqnarray}
\Theta_3(z; \tau_1, \tau_2) \equiv \sum_{s=1}^{n} e^{2\pi i \left( \frac{1}{4} \tau_2 s^4 + \frac{1}{2} \tau_1 s^2 + zs \right)/n},
\end{eqnarray}
is inevitable to know.  This is surely related to the quartic residue problem of integers.  

At this point we are informed that the mathematical study on the generalized Gauss sum and its relation to the elliptic function (Weierstrass $\wp$ function) are well developed recently in number theory \cite{generalized Gauss sum}. 

Now, our problem of quantization begins to touch on the delicate beauty of the number theory.  Therefore, we are better to stop our primitive discussion here, and leave the more careful analysis to another occasion in the future.  It should be pointed out also that the $p$-adic quantum mechanics already takes into account the delicacy of $p$-adic numbers, and utilizes the Gauss sum and the Jacobi sum, and hence there are many things to be referred in this field \cite{p-adic QM}.

\section*{Acknowledgements}
The authors give their thanks to Ken Yokoyama and Shiro Komata for useful discussion and comment on $p$-adic quantum mechanics and Galois theory.  They are grateful to So Katagiri for useful information on the recent development in number theory.  

One of the author (M.S.) are grateful to Hisayoshi Senshiki and Masafumi Sougawa for their understanding and encouragement.


\end{document}